\documentclass[11pt]{article}

\usepackage{color}
\definecolor{indigo}{RGB}{0,0,120}
\usepackage[colorlinks=true, linkcolor=indigo, citecolor=blue, urlcolor=indigo]{hyperref}
\usepackage[margin=1in]{geometry}
\usepackage{amsmath}
\usepackage{amssymb}
\usepackage{comment}

\usepackage{graphicx}
\usepackage{amsfonts}
\usepackage{mdframed}

\usepackage{caption}
\usepackage{subcaption}

\newcommand{\beq}{\begin{equation}}
\newcommand{\eeq}{\end{equation}}
\newcommand{\beqs}{\begin{eqnarray}}
\newcommand{\eeqs}{\end{eqnarray}}
\newcommand{\half}{\frac{1}{2}}
\newcommand{\dd}[2]{\frac {\partial #1}{\partial #2}}
\newcommand{\DD}[2]{\frac {d #1}{d #2}}
\newcommand{\pdr}{\partial}
\newcommand{\ov}[1]{\frac{1}{#1}}
\newcommand{\lambdabar}{{\mkern0.75mu\mathchar '26\mkern -9.75mu\lambda}}

\newcounter{rowno}
\newcommand{\bra}{\langle}
\newcommand{\ket}{\rangle}

\newcommand{\grad}{\nabla}

\def\del{\delta}			 
			 
		\def\sig{\sigma}		
		\def\tht{\theta}

\newcount\colveccount  
\newcommand*\colvec[1]{\global\colveccount#1  \begin{pmatrix} \colvecnext} \def\colvecnext#1{#1 \global\advance\colveccount-1
        \ifnum\colveccount>0 \\ \expandafter\colvecnext
        \else \end{pmatrix} \fi}

\newcommand{\bfv}{{\bf v}}
\newcommand{\bfq}{{\bf q}}
\newcommand{\bfw}{{\bf w}}

\newcommand{\bfF}{{\bf F}}
\newcommand{\bfPhi}{{\bf \Phi}}
\newcommand{\bfU}{{\bf U}}
\newcommand{\bfr}{{\bf r}}

\newcommand{\bfhatn}{{\bf \hat n}}

\newcommand{\bfG}{{\bf G}}
\newcommand{\bfg}{{\bf g}}

\newcommand{\calR}{{\mathcal R}}

\title{The Added Mass Effect and the Higgs Mechanism}
\author{Govind S. Krishnaswami and Sachin Phatak \\ {\small Chennai Mathematical Institute, H1, SIPCOT IT Park, Siruseri 603103, India}\\{\small Email: govind@cmi.ac.in, phatak@cmi.ac.in}}
\date{May 10, 2020\\\vspace{0.5cm}
\small Published by the Indian Academy of Sciences in \href{https://doi.org/10.1007/s12045-020-0936-8}{Resonance {\bf 25}(2), 191 (2020).}
}

\begin{document}

\maketitle

\scriptsize

\begin{abstract}

A rigid body accelerated through an inviscid, incompressible fluid appears to gain mass, which is encoded in an added mass tensor. Swimmers, air bubbles, submarines and airships are slowed down by the associated `added mass' force proportional to their acceleration, which is distinct from viscous drag and buoyancy. In particle physics, otherwise massless electrons, quarks, $W$ and $Z$ bosons, moving through the Higgs scalar field acquire masses encoded in a mass matrix. In this expository article we give an introduction to the fluid mechanical added mass effect through examples of potential flows in various dimensions and exploit a correspondence with the Higgs mechanism [introduced in \href{https://doi.org/10.1098/rspa.2014.0803}{Proc. R. Soc. A {\bf 471}: 20140803 (2015)}] to relate this to how the $W$ and $Z$ bosons can get their masses, while leaving the photon massless. The correspondence relates the Higgs scalar field to the fluid, the vacuum expectation value of the Higgs field to a constant fluid density and quantum fluctuations around the Higgs condensate to compressional waves in the fluid. The shape of the rigid body encodes the pattern of gauge symmetry breaking through the eigenvalues of the corresponding mass matrices. Possible directions of acceleration of the rigid body are related to directions in the gauge Lie algebra with a `flat' direction in the body corresponding to a massless photon. Moreover, symmetries of the body are related to those of the scalar vacuum manifold. The Higgs boson may be viewed as the analog of a long wavelength fluid mode around an accelerated body.\\

\end{abstract}


\normalsize

\section{Introduction}
\label{sec-intro}

It turns out that air bubbles would rise 400 times faster in water if buoyancy were the only force acting on them. Submarines and airships must carry more fuel than one might expect even after accounting for viscous effects and form drag\footnote{Form drag has to do with loss of energy to infinity: waves can propagate and carry energy to infinity even in a flow without viscosity.}. In fact, one would have to apply a larger force while playing volleyball under water than in air or in outer space. Air bubbles, submarines, airships and volleyballs appear to have a larger inertia when they are accelerated through a fluid due to the so-called added mass effect. It arises because in order to accelerate a body through an otherwise stationary fluid, one must also accelerate some of the surrounding fluid. On the other hand, no such additional force is required to move a body at {\it constant} velocity through an ideal (incompressible, inviscid and irrotational or vorticity-free) fluid.

The added mass effect is quite different from viscous drag. The former is a frictionless effect that gives rise to an opposing force proportional to the {\it acceleration} of the body, thus adding to the mass or inertia $m$ of the body. This additional inertia $\mu$ is called its added or virtual mass. The added mass depends on the shape of the body, the direction of acceleration relative to the body as well as on the density $\rho$ of the surrounding fluid. For example, the added mass of a sphere is one-half the mass of fluid displaced by it. Unlike the moment of inertia of a rigid body, its added mass does not depend on the distribution of mass within the body; in fact it is independent of the mass of the body, but grows with the density of the fluid. The more familiar viscous drag on a body is an opposing frictional force that depends on its \emph{speed}: for slow motion it is proportional to the speed, but at high velocities it can be proportional to the square of the speed. The added mass effect is also different from buoyancy. The latter always opposes gravity and unlike the added mass effect, is present even when the body is stationary.

The added mass effect was identified in the early 1800s through the work of Friedrich Wilhelm Bessel, George Gabriel Stokes, Sim\'eon Denis Poisson and others. In his 1850 paper \emph{On the effect of the internal friction of fluids on the motion of pendulums} \cite{ref-stokes}, Stokes is mainly concerned with, ``the correction usually termed the reduction to a vacuum'' of a pendulum swinging in air. He credits Bessel with the discovery of an additional effect, over and above buoyancy, which appears to alter the inertia of the pendulum swinging in air. According to Bessel, this added mass was proportional to the mass of the fluid displaced by the body. Stokes says, ``Bessel represented the increase of inertia by that of a mass equal to $k$ times the mass of the fluid displaced, which must be supposed to be added to the inertia of the body itself.'' In this same paper, Stokes also refers to Colonel Sabine, who directly measured the effect of air by comparing the motion of a pendulum in air with that in a large vacuum chamber. The added inertia for the pendulum in air was deduced to be $0.655$ times the mass of air displaced by the pendulum. Stokes attributes the first mathematical derivation of the added mass of a sphere to Poisson, who discovered that the mass of a swinging pendulum is augmented by half the mass of displaced fluid.

The added mass effect may be understood through the simplest of ideal flows. Consider purely translational motion\footnote{It could be a challenging task for an external agent to ensure that an irregularly shaped body executes purely translational motion, i.e. to ensure that there are no unbalanced torques about its centre of mass that cause the body to rotate.} of a rigid body of mass $m$ in an inviscid, incompressible and irrotational fluid at rest in 3-dimensional space. To impart an acceleration $\bf a = \dot {\bf U}$ to it, an external agent must apply a force $\bfF^{\rm ext}$ exceeding $m {\bf a}$. Newton's second law relates the components of this force to those of its acceleration:
	\beq
	 F_i^{\rm ext} = m \: a_i + F^{\rm add}_i \quad \text{where} \quad i =1,2,3.
	\eeq
Here $F^{\rm add}_i = \sum_{j=1}^3 \mu_{ij} \: a_j$. Part of the externally applied force goes into producing a fluid flow. The added force $\bfF^{\rm add}$ is proportional to acceleration, but could point in a different direction, depending on the shape of the body. The constant $3 \times 3$ symmetric matrix $\mu_{ij}$ that relates acceleration to the added force is the `added-mass tensor'. Unlike the inertia tensor, $\mu_{ij}$ is independent of the distribution of mass in the body, though it depends on the fluid and the shape of the body. The added mass tensor $\mu_{ij}$ for a sphere is a multiple of the identity matrix: $\mu_{ij} = \mu \del_{ij}$. In other words, the added mass $\mu$ of a sphere is the same in all directions. When a sphere is accelerated horizontally in water, it feels a horizontal opposing acceleration-reaction force ${\bf G} = - {\bf F}^{\rm add} = - \mu {\bf a}$ aside from an upward buoyant force, an opposing viscous force etc. It turns out that the added mass grows roughly with the cross sectional area presented by the accelerated body. For instance, a flat plate has no added mass when accelerated along its plane.

In this article, we explain these ideas and introduce the added mass effect through examples. We show how the added mass tensor may be calculated for one, two and three-dimensional flows and discuss some of its features and consequences. In the final Section, we describe a surprising connection to the Higgs mechanism in particle physics \cite{ref-thyagaraja}. It turns out that the force carriers of the weak interactions ($W$ and $Z$ bosons) as well as the basic matter particles (quarks and leptons) are, strictly speaking, massless. However, they gain a mass through their interactions with a `condensate' of the Higgs field\footnote{The Higgs mechanism does {\it not} explain the masses of certain other particles such as the proton and the neutron. The latter are a lot heavier ($\approx938$ MeV$/c^2$) than the sum of the masses ($\approx 2$-$5$ MeV$/c^2$) of their three valence quarks. The binding energy of gluons is believed to contribute significantly to the proton and neutron masses.}. We use an analogy \cite{ref-ham} with the fluid mechanical added mass effect to intuitively explain some features of the Higgs mechanism in particle physics. This `Higgs Added Mass' correspondence is not quite a mathematical duality like Kramers-Wanniers \cite{kramers-wannier} or AdS/CFT \cite{maldacena} but a physical analogy that provides a new viewpoint on both subjects.

\section{One-dimensional flow along a circle}

\begin{figure}
 \centering
 \includegraphics[scale=0.3]{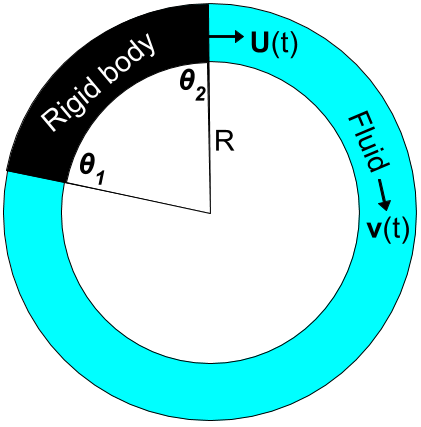} 
 \caption{Arc-shaped rigid body accelerated through fluid along a circle} \label{fig-arc-in-circular-fluid}
 \end{figure}

We begin by introducing the added mass effect through a simple example of incompressible flow in one dimension. Incompressible here means the fluid always has the same density everywhere. Consequently, the velocity $v$ of the fluid must be the same everywhere, though it could depend on time. Suppose an arc-shaped rigid body of length $L$ moves along the rim of a circular channel of radius $R$ (see Fig.~\ref{fig-arc-in-circular-fluid}). We will suppose that the ends of the body are at the angular positions $\theta_1(t)$ and $\theta_2(t)$ so that $R (\theta_2 - \theta_1) = L$. The fluid occupying the rest of the circumference of the channel has velocity $v(t)$ tangent to the circle at all angular positions $\theta$.

Now imagine an external agent moving the body at speed $U(t)$ so that its ends have the common speed $R \dot \tht_1 = R \dot \tht_2 = U(t)$. Since the fluid cannot enter the body, it must have the same speed as the end-points of the body: $v(\tht_1,t) = v(\tht_2,t) = U(t)$. Thus, the fluid instantaneously acquires the velocity of the body everywhere: $v(t) = U(t)$. Signals can be communicated instantaneously since the speed of sound is infinite in an incompressible flow.

In the absence of the fluid, an external agent would have to supply a force $M \dot U$ to accelerate the body. To find the additional force required in the presence of the fluid, we consider the kinetic energy of the fluid:
\beq
	E = \half \int_{\rm fluid} \rho \: v^2(t) \:  R \, d \tht = \half \rho v^2(t) (2\pi R - L).
\eeq
The rate of change of flow kinetic energy $\dot E$ is the rate at which energy must be pumped into the flow:
\beq
	\dot E = \rho \: v(t) \: \dot v(t) \: (2\pi R - L) = \rho \: U(t) \: \dot U(t) \: (2\pi R - L).
\eeq
$\dot E$ must equal the extra power supplied by the external agent, i.e., $\dot E = F^{\rm add} U(t)$. Thus, the additional force
\beq
	F^{\rm add} = \rho \: \dot U(t) \: (2\pi R - L)
\eeq
is proportional to the body's acceleration. $G = -F^{\rm add}$ is called the `acceleration reaction force'. The constant of proportionality
\beq
	\mu = \rho(2 \pi R - L)
\eeq
is called the added mass. Notice that $\mu$ is equal to the mass of fluid. This is peculiar to flows in one dimension. Had we taken the fluid to occupy an infinitely long channel (instead of a circle), the added mass would have been infinite. Furthermore, the added mass is proportional to the fluid density $\rho$ and depends on the shape of the body, but is independent of the body's mass. Crucially, the added force is proportional to the body's acceleration as opposed to its velocity. In particular, a body moving uniformly would not acquire an added mass.\footnote{The result that uniformly moving finite bodies in an unbounded steady potential flow (see Section \ref{sec-2d-flow-cylinder}) do not feel any opposing force is a peculiarity of inviscid hydrodynamics called the d'Alembert paradox. Strictly speaking, this result is valid only in the absence of `vortex sheets' and `free streamlines'. In commonly encountered fluids, viscous forces introduce dissipation and surface waves carry away energy to `infinity' so that an external force is required even to move a body at constant velocity.}

We next turn to bodies accelerated through two- and three-dimensional flows, which are much richer than the one-dimensional example above. An intrepid reader who cannot wait to explore the analogy with the Higgs mechanism may proceed directly to Section \ref{sec-analogy}.

\section{Two-dimensional flow around a cylinder}
\label{sec-2d-flow-cylinder}

\begin{figure}
 \centering
 \includegraphics[scale=0.4]{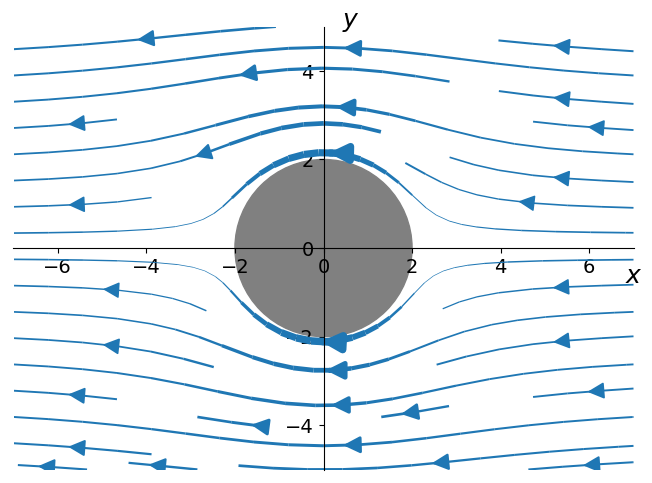} 
 \caption{Snapshot of velocity field in the $x$-$y$ plane for potential flow past a stationary right circular cylinder of radius $a = 2$ with axis along $z$. Asymptotically $\bfv \to -\hat x$ ($U = 1$). The thickness of flow lines grows with fluid speed.} \label{fig-flow-around-cylinder}
 \end{figure}

We next consider inviscid flow perpendicular to the axis of an infinitely long right circular cylinder of radius $a$ with axis along $\hat z$, as shown in Fig.~\ref{fig-flow-around-cylinder}. We will find the added mass per unit length of the cylinder by determining the velocity field of the fluid flowing around it.

Although the fluid moves in 3d space, the flow is assumed to be quasi two-dimensional with translation invariance in the $z$-direction. Thus, we take the $z$-component of the flow velocity to vanish so that $\bfv$ points in the $x$-$y$ plane. For simplicity, we further take the flow to be irrotational ($\grad \times \bfv = 0$) which allows us to write $\bfv = \grad \phi$ in terms of a velocity potential $\phi$. We also take the flow to be incompressible ($\grad \cdot \bfv = 0$, this is reasonable as long as the flow speed is much less that that of sound) which requires the velocity potential to satisfy Laplace's equation $\grad^2 \phi = 0$. To find $\phi$, we must supplement Laplace's equation with boundary conditions. The fluid cannot enter the body, so $\bfhatn \cdot \bfv = \bfhatn \cdot \grad \phi = 0$ on the surface of the body. Here $\bfhatn$ is the outward-pointing unit normal on the body's surface. Additionally, we suppose that far away from the body the fluid moves uniformly: $\bfv \rightarrow -U \hat x$. In other words, the cylinder is at rest while the fluid moves leftward past it.

Of course, we are interested in a cylinder accelerating through an otherwise stationary fluid. However, it is easier to solve Laplace's equation in a region with fixed boundaries. So, we begin by considering flow around a stationary cylinder. After finding the velocity field of this flow, we will apply a Galilean boost and move to a frame where the cylinder moves at velocity $U \hat x$. By making $U$ time-dependent, we will find the acceleration-reaction force and added mass of the cylinder.

Our first task is to solve Laplace's equation in the $x$-$y$ plane subject to the impenetrability boundary condition ${\bfhatn \cdot \grad \phi} = \frac{\pdr \phi}{\pdr r} = 0$ at $r = a$ and the asymptotic condition $\phi \to -U r \cos \theta$ as $r \to \infty$ (so that $\bfv \to -U\hat x$ as $r \to \infty$). Here, the cylinder is assumed centered at the origin about which the plane-polar coordinates $r = \sqrt{x^2 + y^2}$ and $\theta = \tan^{-1}(y/x)$ are defined. Laplace's equation in these coordinates takes the form
\beq
	\grad^2 \phi = \frac{1}{r}\frac{\pdr}{\pdr r} \left( r \frac{\pdr \phi}{\pdr r} \right) + \frac{1}{r^2} \frac{\pdr^2 \phi}{\pdr\theta^2} = 0.
	\label{eqn-laplace-cylinder}
\eeq
We will solve this linear equation by separation of variables and the superposition principle. Let us suppose that $\phi$ is a product $\phi(r, \theta) = R(r) \Theta(\theta)$ where $\Theta(\theta + 2\pi) = \Theta(\theta)$ is periodic around the cylinder. Upon division by $R \,\Theta$ the partial differential equation (\ref{eqn-laplace-cylinder}) becomes a pair of ordinary differential equations. Indeed, we must have
\beq
	\ov{R} r \frac{d}{dr} \left(r \frac{dR}{dr}  \right) = -\ov{\Theta} \DD{^2\Theta}{\theta^2} = n^2 = {\rm constant}.
\eeq
The separation constant $n^2$ must be positive for $\Theta$ to be periodic. In fact, the equation $\Theta'' = -n^2 \Theta$ describes a simple harmonic oscillator with solutions
\beq
	\Theta_n(\theta) = A_n \cos n\theta + B_n \sin n\theta.
	\label{eqn-theta-periodicity-cylinder}
\eeq
Periodicity and linear independence then require $n$ to be a non-negative integer. For each such $n$, the radial equation has the solution
\beq
	R_n(r) =
	\begin{cases}
		C_n r^n + \frac{D_n}{r^n} 	& {\rm for} \;\; n > 0 \;\; {\rm and}\\
		C_0 + D_0 \ln r 			& {\rm for} \;\; n = 0.		
	\end{cases}
\eeq
Here $A_n, B_n, C_n$ and $D_n$ are constants of integration. By the superposition principle, the general solution of (\ref{eqn-laplace-cylinder}) is
\beq
	\phi(r, \theta) = (C_0 + D_0 \ln r) + \sum_{n=1}^\infty \left( C_n r^n + \frac{D_n}{r^n} \right) \left( A_n \cos n\theta + B_n \sin n\theta \right).
\eeq
The asymptotic boundary condition implies that $C_0 = D_0 = 0$, $C_1 = -U$ and $A_1 = 1$ while $C_n = A_n = B_n = 0$ for all other $n$. The impenetrability condition gives $D_1 = -U a^2$. Thus, the velocity potential for flow around the cylinder is
\beq
	\phi(r, \theta) = -U \cos\theta \left(r + \frac{a^2}{r} \right).
\eeq
The corresponding velocity field is 
	\beq
	\bfv = \grad \phi = -U \hat x + U \frac{a^2}{r^2} \left( \cos \theta\, \hat r + \sin \theta \,\hat \theta \right).
	\eeq

Now we make a Galilean boost to a frame where the fluid is stationary at infinity, but the cylinder moves rightward with velocity $U \hat x$. The resulting velocity field around the moving cylinder is
\beq
	\bfv' = \bfv + U \hat x = U \frac{a^2}{r'^2} \left( \cos \theta'\, \hat r' + \sin \theta' \, \hat \theta' \right),
\eeq
where $r'$ and $\theta'$ are defined relative to the instantaneous centre of the cylinder.

To investigate the added mass effect, we suppose an external agent wishes to accelerate the cylinder (of mass $M$ per unit length) at the rate $\dot {\bf U} = \dot U \hat x$. Part of the energy supplied goes into the kinetic energy of the flow and is manifested as the added mass of the cylinder. Just as the kinetic energy of the cylinder per unit length ($\half M U^2$) is quadratic in $U$, so is that of the flow:
\beq
	K_{\rm flow} = \half \rho \int_0^{2\pi}\int_a^\infty \left(\bfv'\right)^2 r' \, dr'\, d\theta' = \half \rho \iint \frac{U^2 a^4}{r'^3}dr'\,d\theta' = \half \rho \pi a^2 U^2 \equiv \half \mu U^2.
\eeq
 Here $\rho$ is the density of the fluid per unit area in the $r,\theta$ plane. Thus the total kinetic energy per unit length supplied by the agent is
\beq
	K_{\rm total} = \half \left( M + \mu \right) U^2.
\eeq
The associated power supplied is $\dot K_{\rm total} = (M + \mu) \dot {\bf U} \cdot {\bf U} \equiv {\bf F^{\rm ext}} \cdot {\bf U}$. Thus, a force ${\bf F^{\rm ext}} = \left( M + \mu \right) \dot {\bf U}$ is required to accelerate the body at ${\bf\dot U}$. The mass of the cylinder therefore appears to be augmented by an added mass per unit length $\mu = \rho \pi a^2$. We notice that the added mass of the cylinder is equal to the mass of fluid displaced, though this is not always the case as we will learn in Section \ref{sec-derivation}.

More generally, it can be shown that if the cylinder had an elliptical rather than circular cross section, the added mass is different for acceleration along the two semi-axes. If $a_1$ and $a_2$ are the lengths of the two semi-axes, the added masses are $\mu_{1,2} = \rho \pi a_{2,1}^2$ for motion along the corresponding semi-axis. Thus, the added mass is smaller when the cylinder presents a smaller cross section.

\section{Added mass tensor in three dimensions}
\label{sec-derivation}

We have seen that to accelerate a cylinder of mass $m$ perpendicular to its axis in an ideal fluid, an external agent must provide a force $\bfF^{\rm add} = \mu \dot \bfU$ in addition to the inertial force $m \dot \bfU$ where $\bfU$ is the velocity of the cylinder. More generally, the body's acceleration need not be directed along ${\bf F}^{\rm add}$, though we will see that the two are linearly related: $F^{\rm add}_i = \sum_{j=1}^3 \mu_{ij} \dot U_j$. For flows in 3 dimensions, the added mass tensor $\mu_{ij}$ is a real, symmetric, $3 \times 3$ matrix with positive eigenvalues. It turns out that $\mu_{ij}$ is proportional to the (constant) density $\rho$ of the fluid and depends on the shape of the rigid body. However, unlike the inertia tensor\footnote{The inertia tensor of a rigid body is the $3\times 3$ matrix $I_{ij} = \int \rho(\bfr) \, (r^2 \del_{ij} - r_i r_j) \, d^3r$ where $\rho(\bfr)$ is its mass density and the integral extends over points $\bfr$ in the rigid body.} $I_{ij}$ of a rigid body, $\mu_{ij}$ is independent of its mass distribution. For example, for a sphere of radius $a$, the added mass tensor is a multiple of the identity, $\mu_{ij} = \frac{2}{3}\pi a^3 \rho\, \del_{ij}$. In other words, the added mass of a sphere is half the mass of fluid displaced irrespective of the direction of acceleration which is always along $\bfF^{\rm add}$. In particular, for an air bubble in water, the added mass is about $\half \rho_{\rm water}/\rho_{\rm air} \approx 400$ times its actual mass as indicated in Section \ref{sec-intro}. More generally, for bodies less symmetrical than a sphere, the added mass tensor need not be a multiple of the identity and the added masses along different `principal' directions can be different.

Let us illustrate a simple consequence of the added mass being a tensor rather than a scalar. Suppose our rigid body is irregularly shaped and has an added mass tensor $\mu_{ij}$ with non-zero off-diagonal entries. In order to accelerate it along the $\hat x$ direction, we must apply a force $\bfF = m\dot U \hat x + \bfF^{\rm add}$, where $\bfF^{\rm add} = \dot U \left( \mu_{11}\hat x + \mu_{21} \hat y + \mu_{31} \hat z\right)$. Thus, to accelerate the body along $\hat x$, we would have to supply an added force in a different direction. On the other hand, even an irregularly shaped rigid body always has (at least) three `principal axes'. They have the property that the added mass tensor is diagonal $({\rm diag}(\mu_1, \mu_2, \mu_3))$ when expressed in the principal axis basis. Thus, for instance, a force along the second principal axis produces an acceleration in the {\em same} direction with added mass $\mu_2$. The principal axes are the eigenvectors of $\mu_{ij}$ and $\mu_{1,2,3}$ are the corresponding eigenvalues.

In this Section we derive the added mass effect in 3-dimensional flows and express $\mu_{ij}$ as an integral over the surface of the body \cite{ref-batchelor}. As before, consider inviscid, incompressible and irrotational flow around a rigid body (assumed to be simply connected) in a large container. For simplicity, we assume that the external agent accelerates the body along a straight line without rotating it. The fluid is assumed to be at rest far from the body ($\bfv \to 0$ as $|\bfr| \to \infty$) and its velocity ${\bf v}$ is expressed in terms of a potential $\bfv = \grad \phi$. Due to incompressibility, $\phi$ must satisfy Laplace's equation $\grad \cdot \bfv = \grad^2 \phi = 0$. Impenetrability requires the boundary condition (BC) $\grad \phi \cdot \bfhatn = {\bf U}(t) \cdot \bfhatn$ on the body's surface where $\bfhatn$ is the unit outward normal.

The information in $\phi$ may be conveniently packaged in a `potential vector field' $\Phi(\bfr, t)$. To see this, notice that Laplace's equation and the BC is a system of inhomogeneous linear equations of the form $L \phi = b$ where $b$ is linear in $\bfU$, with solution $\phi = L^{-1}b$. Thus, $\phi$ must be linear in $\bfU$ and may be expressed as $\phi = \bfU \cdot \Phi$. Here $\Phi(\bfr, t)$ is independent of $\bfU$ and can depend only on the position of the observation point $\bfr$ relative to the body's surface. Being rigid and in rectilinear motion, the surface of the body is determined by the location of a marked point in the body, which may be chosen say, as the center of volume ${\bf r}_0$. Thus, 
\beq
\phi(\bfr, t) = {\bf U(t)} \cdot {\bf \Phi}({\bf r} - {\bf r}_0(t)).
\eeq
For example, for a sphere of radius $a$ centered at the origin at time $t_0$,
\beq
\bfPhi({\bf r}, t_0) = -\frac{1}{2} a^3 \frac{\bf \hat{r}}{r^2} \;\; {\rm and} \;\; \phi(\bfr, t_0) = -\frac{1}{2} a^3 \bfU \cdot \frac{\bf \hat{r}}{r^2}.
\eeq

\subsection{Finding the added mass tensor from the fluid pressure on the body}
\label{sec-derivation-subsec-1}

To obtain the pressure force on the body, we use a generalization of Bernoulli's equation $(p + \half \rho v^2 = {\rm constant})$ to time-dependent potential flows with constant density:
\beq
p + \half \rho v^2 + \rho \dd{\phi}{t} \: = \: B(t)
\label{e-bernoulli-eqn}
\eeq
where $B(t)$ is a function of time alone. This version of Bernoulli's equation is derived in Appendix \ref{a:bernoulli-unsteady-potential-flow} and may be used to write the force due to pressure on the body as an integral over its surface $A$:
\beq
{\bf F} \; = \; - \int_{A} p \, {\bfhatn} \, dA = \rho \int_{A} \left( \frac{\partial \phi }{\partial t} + \frac{1}{2} v^2 \right) {\bfhatn} \, dA.
\eeq
The Bernoulli constant $B(t)$ does not contribute as the integral $\int_A \bfhatn \, dA$ vanishes over the closed surface $A$ of the body. 
We may write $\bfF$ as a combination of the acceleration reaction force $\bfG$ and a $\bfG'$ which does not depend on acceleration, by using the factorization $\phi = {\bf U} \cdot {\bf \Phi}$:
\beq
	{\bf F} = \rho \int_{A} {\bf\dot{U}} \cdot {\bf \Phi} \: {\bfhatn} \: dA
+ \int_{A} \left[ \half \rho v^2 - \rho {\bf U} \cdot {\bf v} \right] \: {\bfhatn} \: dA \equiv \bfG + \bfG'.
\eeq
In fluids which are at rest at infinity in $\mathbb{R}^3$ the second integral $\bfG'$ vanishes \cite{ref-batchelor}. In a large flow domain of linear extent $R$, a multipole series\footnote{For large $r = |\bfr|$, $\phi$ admits the multipole expansion: $\phi = c_0 + c_1/r + c_2/r^2 + \cdots$. $c_0$ can be chosen to vanish as it does not affect the velocity. The monopole coefficient $c_1$ must vanish since there are no sources or sinks in the fluid, just as it does for the electrostatic potential around a neutral body. Therefore, $\phi$ can be at most of order $1/r^2$ asymptotically. Consequently, $|\bfv| = |\grad\phi| \sim 1/r^3$. This can be used to show that the non-acceleration reaction force $\bfG'$ vanishes if the fluid extends to infinity in all directions.} for $\phi$ may be used to show that $\bfG'$ is at most of ${\cal O}(1/R)$. It is as though the liquid can bounce off the boundary of the domain and come back to hit the body; this effect will be ignored henceforth. Though we do not discuss it here, upon accounting for the effect of gravity (via the acceleration due to gravity $\bfg$), $\bfG'$ includes a buoyancy contribution $- \rho {\rm Vol}_{\rm body} \, {\bfg}$ equal to the weight of displaced fluid. Thus, we may write the acceleration reaction force as
\begin{equation}
G_i = -F_i^{\rm add} = - \mu_{ij} \dot U_j \quad \textmd{where}\quad 
\mu_{ij} = - \rho \int_{A} \Phi_j \: n_i \: dA.
\end{equation}
$\mu_{ij}$ is the added-mass tensor, it is a direction-weighted body-surface average of the potential vector field $\Phi$. It can be shown to be symmetric and independent of time, though it does depend on the shape of the body and is proportional to the density of the fluid. Now, the rate at which energy is added to the flow is
\beq
{\bf F^{\rm add}} \cdot {\bf U}(t) = \sum_{i,j=1}^3\mu_{ij} \dot U_j U_i = \DD{}{t}\left(\sum_{i,j=1}^3\half \mu_{ij} U_i U_j\right).
\eeq
Hence, the kinetic energy $K$ of the flow can be written\footnote{See \cite{kambe} for an alternate approach to computing the flow energy.} exclusively in terms of the added mass tensor and the velocity of the body:
\beq
K = \half \int_V \rho v^2 \: dV \; = \; \sum_{i,j=1}^3\frac{1}{2}\mu_{ij} U_i U_j.
\eeq
Since $K \geq 0$, $\mu_{ij}$ must be a positive matrix, i.e., one with non-negative eigenvalues.

\subsection{Examples of added mass tensors}

\begin{figure}
	\centering
	\includegraphics[scale=0.4]{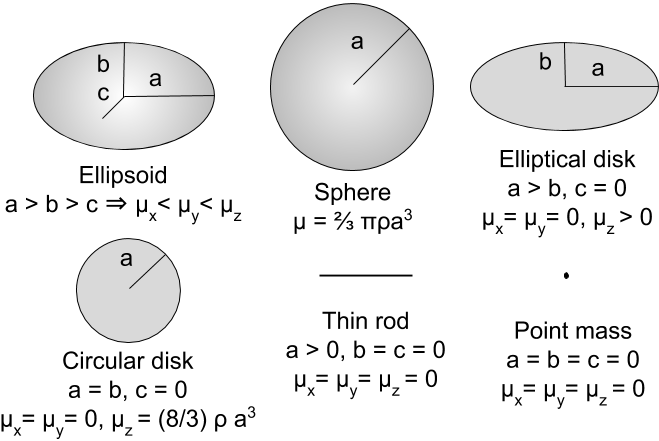}
	\caption{Examples of rigid bodies (an ellipsoid and its limiting cases) and their principal added masses.} \label{fig-bodies-and-added-masses}
\end{figure}

By solving for potential flow around various rigid bodies, one obtains their added-mass tensors (see \cite{ref-marine-hydrodynamics, ref-lamb} and Fig.~\ref{fig-bodies-and-added-masses}). For example, the added mass of a sphere with radius $a$ is one-half the mass of displaced fluid independent of the direction of acceleration: $\mu_{ij} = \frac{2}{3} \pi a^3 \rho \del_{ij}$. 

For an ellipsoid $(\frac{x^2}{a^2} + \frac{y^2}{b^2} + \frac{z^2}{c^2} =  1)$ with semi-axes $a, b$ and $c$, $\mu_{ij}$ becomes diagonal in the principal axis basis. If the semi-axes are ordered according to $a > b > c$, then the eigenvalues corresponding to the $x, y$ and $z$ principal axes may be shown to satisfy $\mu_x < \mu_y < \mu_z$. Roughly, the added-mass grows with cross-sectional area presented. In particular, if $a = b$ (ellipsoid of revolution), the corresponding added-mass eigenvalues $\mu_x$ and $\mu_y$ are equal. Taking $c \to 0$, the ellipsoid becomes an elliptical disk with vanishing added mass eigenvalues for acceleration along its plane, in which case the body does not present any cross-sectional area: $\mu_x = \mu_y = 0$. Keeping $c = 0$ and taking $a=b$, we get a circular disk whose principal added-masses are $\mu_x = \mu_y = 0$ and $\mu_z = (8/3)\rho a^3$. Shrinking an elliptical disk further ($c \to 0$ and $b \to 0$), we find that a slender rod of length $2a$ has vanishing added-mass for acceleration along any direction. Subject to impenetrable boundary conditions, such a rod cannot displace fluid. Similarly, a point particle and a one-dimensional body (e.g. a bent wire) moving in three dimensions have vanishing added masses.

The virtual inertia per unit length of an infinite right circular cylinder, when accelerated perpendicular to its axis, equals the mass of displaced fluid (see Section \ref{sec-2d-flow-cylinder}). In fact, the added mass tensor per unit length of a cylinder of radius $a$ with axis along $z$, is given by $\mu_{ij}/L = \pi a^2 \rho\; {\rm diag}(1,1,0)$.

More generally, the added mass effect extends to two-dimensional as well as four and higher dimensional flows. For example, a circular disk of radius $a$ accelerated through planar potential flow has a $2 \times 2$ added mass matrix $\mu_{ij} = \pi \sig a^2 \del_{ij}$. Here $\sig$ is the (constant) areal density of fluid.

\section{Correspondence with the Higgs mechanism}
\label{sec-analogy}

Just as photons mediate the electromagnetic force, the $W^{\pm}$ and $Z^0$ gauge bosons mediate the weak nuclear force responsible for beta decay. The mass of each of these force carriers is inversely proportional to the range of the corresponding force. For instance, the Coulomb potential ($\propto 1/r$) between electric charges has an effectively infinite range corresponding to the masslessness of photons. Similarly the range of the strong nuclear force between protons and neutrons is given by the range of the Yukawa potential ($\propto e^{-r/\lambdabar}/r$). The latter is of order the reduced Compton wavelength $\left(\lambdabar = \hbar/m c \approx 1~{\rm femtometer}\right)$ of $\pi$-mesons $(m \approx 0.139$~GeV$/c^2)$ which mediate the strong force.

On the other hand, the weak nuclear force is very short ranged ($\approx 0.001 ~{\rm femtometers}$), which requires the $W^{\pm}$ and $Z^0$ gauge bosons to be very massive ($80$-$91$GeV$/c^2$; by contrast, the proton mass is only $0.938$GeV$/c^2$). However, including mass terms for the $W^{\pm}$ and $Z^0$ particles in the theory explicitly breaks the `gauge' symmetry of the equations and unfortunately destroys the predictive power (`renormalizability') of the theory. The Higgs mechanism solves this problem by a process of `spontaneous' rather than `explicit' breaking of gauge symmetry.\footnote{As mentioned earlier, while the Higgs mechanism does explain the masses of the $W$ and $Z$ bosons as well as the quarks (e.g. up and down) and leptons such as the electron and muon, it is not relevant to explaining why the proton or neutron are so much more massive than the quarks that make them up. The latter has to do with the binding energy of gluons.} It was proposed in the work of several physicists including Peter Higgs \cite{higgs}, Robert Brout, Fran\c{c}ois Englert \cite{englert}, Gerald Guralnik, Carl Hagen and Tom Kibble \cite{kibble, guralnik} in 1964, building on earlier work of Philip Anderson \cite{anderson} in superconductivity (see Box 1). The $W^{\pm}$ and $Z^0$ bosons are nominally massless, but appear to be massive due to interactions with the Higgs scalar field whose condensate permeates all of space like a fluid (the strength of this condensate is measured by the vacuum expectation value (vev) of the Higgs field $\approx 246$GeV$/c^2$). Thus gauge symmetry is not spoilt and the predictive power of the theory is restored. It turns out that in the Higgs mechanism the mass terms arise through a $4 \times 4$ matrix whose eigenvalues are the masses of the $W^{\pm}$ and $Z^0$ gauge bosons as well as the photon.

\begin{center}
	\begin{mdframed}
		\begin{center}
			{\bf Box 1: Non-zero photon mass in superconductors and plasmas}
		\end{center}
		Photons in vacuum are massless and travel at the speed $c$ of electromagnetic waves or light. Each component of the electric and magnetic field in an EM wave satisfies the d'Alembert wave equation $\ov{c^2}\frac{\partial^2 \chi}{\partial t^2} - \grad^2 \chi = 0$. These waves are transversely polarized, since in the absence of charges ${\grad \cdot {\bf E}} = \grad \cdot {\bf B} = 0$, so that these fields (${\bf E}(\bfr, t) = {\bf E}_{\bf k} e^{i ({\bf k}\cdot \bfr - \omega t)}$ where $\omega = c |{\bf k}|$) are orthogonal to the direction ${\bf \hat k}$ of propagation (${\bf \hat k} \cdot {\bf E}_{\bf k} = {\bf \hat k} \cdot {\bf B}_{\bf k} =0$). In superconductors and plasmas however, photons become massive - a `mass' term enters the wave equation: $\ov{c^2}\frac{\partial^2 \chi}{\partial t^2} - \grad^2 \chi + m^2 \chi= 0$. Consequently, photons travel at a speed less than $c$ and display both transverse and longitudinal polarizations. This photon mass can be explained by an `abelian' version of the Higgs mechanism. The Meissner effect is a physical manifestation of this photon mass: magnetic fields are expelled from a superconductor except over a thin surface layer whose thickness is given by the London penetration depth. Similarly, in plasmas the electric field of a test charge is screened beyond the Debye screening length \cite{ref-raichoudhury}. Both the penetration depth and screening length are inversely proportional to the photon mass, so that they diverge in vacuum.
	\end{mdframed}
\end{center}
\newpage
\begin{table}[h]
	\small
	\begin{center}
	\begin{tabular}{| p{8cm} | p{8cm} |} \hline
	{\textbf{Added-Mass Effect}} & {\textbf{Higgs Mechanism}} \\ \hline
	 
	Rigid body & Gauge bosons \\ \hline
	Fluid & Scalar field \\ \hline
	Space occupied by fluid & Gauge Lie algebra \\ \hline
	
	Dimension of container & $\dim G = $ number of gauge bosons \\ \hline
	
	Constant fluid density $\rho$ & vev $\bra \phi \ket$ of Higgs scalar condensate \\ \hline
	
	Direction of acceleration of rigid body & Direction in space spanned by gauge bosons \\ \hline
	
	Added-mass tensor $\mu_{ij}$ & Gauge boson mass matrix $M_{ab}$ \\ \hline
	
	Squares of eigenvalues of added mass tensor $\mu_{ij}$ & Eigenvalues of vector boson mass$^2$ matrix \\ \hline
	
	Acceleration along flat face of rigid body & Massless photon \\ \hline
	
	Zero modes of $\mu_{ij}$. E.g. No added mass when a thin plate is accelerated along flat surface. $m = \#$(independent directions along which accelerated motion gives no added mass). & Zero modes of vector boson mass$^2$ matrix. E.g. massless photons in directions of residual gauge symmetry $H$; $m = \dim(H)$. \\ \hline
	
	Spherical rigid body moving in 3d & SU$(2) \to \{ 1 \}$, doublet; equal-mass gauge bosons \\ \hline
	
	Hollow right circular cylindrical shell in 3d & SO$(3) \to$ SO$(2)$, triplet; 2 equal-mass bosons, photon \\ \hline
	
	Ellipsoid of revolution: semi-axes $r_1 = r_2 > r_3$ & System of $W^{\pm}, Z^0$ bosons with $m_{W^{\pm}} < m_Z$ \\ \hline
	
	Acceleration vector of body moving in $d$ dimensions: no added force component $F_e$ if acceleration component $a_e = 0$ where $\hat e$ is an eigen-direction of $\mu_{ij}$. & Vector of gauge couplings $g_i$ for $G = U(1)^d$: Gauge boson $W_1$ has no added mass if coupling $g_1  = 0$.  \\ \hline
	
	Linear dimensions of rigid body & Charges of scalars $n_i$ under $U(1)$ factors of gauge group $U(1)^d$ \\ \hline
	
	3 ways to have zero added mass: acceleration $a_i \to 0$, $\rho \to 0$ and shrink the body to a point. & 3 ways for gauge group to be unbroken: couplings $g_i \to 0$, vev $\bra \phi \ket \to 0$, make scalars uncharged. \\ \hline
	
	Symmetries of curved body & Symmetries of tangent space $T_m ({G/H})$ at point $m$ \\ \hline
	
	Breaking of fore-aft symmetry of pressure distribution on a sphere when accelerated & Gauge symmetry $G$ spontaneously breaks to $H$ when scalars are charged under $G$ and $\langle \phi \rangle \ne 0$.  \\ \hline
	
	`Benign' flow around body moving uniformly through inviscid fluid & Goldstone mode \\ \hline
	
	Different boundary conditions on body surface & Different (non-minimal) couplings between scalars and gauge fields \\ \hline
	
	Newton's law $F_i - ma_i \; = \; \mu_{ij} \; a_j$ & Proca equation $-j^{\nu} + \partial_{\mu}F^{\mu\nu} \; = \; g^2 \eta^2 A^{\nu}$ \\ \hline
	
	Longest wavelength mode in compressible flow around an accelerated body & Higgs boson - longest wavelength mode of Higgs scalar field \\ \hline
	
	Compressional waves in otherwise constant density flow & Quantum fluctuations around constant strength of Higgs condensate \\ \hline
	
	Expansion in powers of Mach number describing effects of compressibility & Semi-classical loop expansion in powers of $\hbar$ \\ \hline

	\end{tabular}
	\normalsize
	
	\caption{Analogies between the added mass effect and the Higgs mechanism.}
	\label{t:HAM}
	
	\end{center}
	
\end{table}
\normalsize

Recalling our discussion of the added mass effect, gauge bosons acquiring masses via the Higgs mechanism sounds like the virtual masses of a rigid body accelerated through a fluid along its principal directions. Moreover, the gauge boson mass matrix is reminiscent of the added mass tensor. In fact, there is a delightful analogy between these two physical phenomena which allows us to intuitively understand the Higgs mechanism by appealing to the fluid mechanical added mass effect \cite{ref-thyagaraja, ref-ham}. The `Higgs added mass correspondence' proceeds as follows. The gauge bosons are collectively like a rigid body and the Higgs field is like the fluid. The density of the fluid is like the strength of the Higgs condensate (the vacuum expectation value of the Higgs field). Moreover, the number of gauge bosons is analogous to the dimension of the space in which the fluid flows. In fact we may picture the correspondence as relating the Euclidean space in which the fluid flows (with the rigid body at its origin) to the linear space spanned by the gauge bosons (the `Lie algebra' of the `gauge group'). Furthermore, a direction for the acceleration of the rigid body is analogous to the choice of a direction in the gauge Lie algebra. As mentioned, the added mass tensor $\mu_{ij}$ plays the role of the gauge boson mass matrix $M_{ab}$. Just as accelerating the body in different directions could result in different added masses, various directions in the Lie algebra could correspond to gauge bosons with possibly different masses. In fact, given a pattern of gauge boson masses, one may associate with it a rigid body moving through an ideal fluid. For instance, if all gauge bosons had equal masses, the corresponding rigid body could be taken to be a sphere. More generally, a rigid body that corresponds to the $W^{\pm}$ and $Z^0$ gauge bosons (with masses $m_{W^+} = m_{W^-} < m_Z$) is an ellipsoid of revolution with semi-axes $r_1 = r_2 > r_3$. Acceleration along the two larger semi-axes (where the body presents a smaller cross section) would result in lesser added masses, corresponding to the lighter $W^{\pm}$ bosons. Interestingly, the acceleration of a rigid body along a flat direction (such as along the plane of a thin sheet) produces no added mass, corresponding to a zero eigenvalue of $\mu_{ij}$. This vanishing added mass is like a massless gauge boson such as the photon and corresponds to the direction of an unbroken gauge symmetry. For example, a hollow right circular cylindrical shell corresponds to a pair of equally massive gauge bosons (for acceleration perpendicular to its axis) and a massless photon (for acceleration along its axis).
\begin{figure}
	\begin{center}
	\includegraphics[scale=0.75]{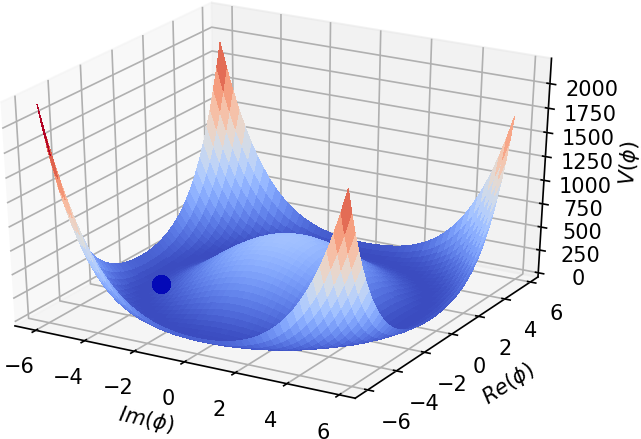}
	\caption{Spontaneous symmetry breaking in the Higgs mechanism is achieved through a potential $V(\varphi) = \lambda (|\varphi|^2 - a^2)$ for the complex scalar field $\varphi$ shown above. Here the vacuum expectation value (vev) of the scalar field is $a = 5$ and the scalar `self-coupling' is $\lambda = 1$. Though $V$ is circularly symmetric, a `particle' (the blue dot) in its minimum energy state must lie at a particular point along the circular valley, thereby `spontaneously' breaking the rotation symmetry. The lowest-lying excitation is a massless `Goldstone mode' corresponding to a particle rolling arbitrarily slowly along the circular valley. The lowest-lying (massive) radial excitation corresponds to the Higgs particle.}
	\label{f:mexican-hat-potential}
	\end{center}
\end{figure}

When a sphere moves at constant velocity through an ideal fluid, its upstream and downstream hemispheres have the same pressure distributions. This fore-aft symmetry breaks when the sphere is accelerated (the pressure on the front hemisphere is on average more than that on its rear). This is somewhat analogous to the spontaneous breaking of the gauge symmetry (see Fig.~\ref{f:mexican-hat-potential}) when the gauge bosons interact with the Higgs field. When a body moves at constant velocity through inviscid flow, it does not require an external force to keep it moving (d'Alembert's paradox) and the flow around it is maintained without any energy input. This `benign' flow is analogous to the `Goldstone mode' which is produced when gauge  bosons do not interact with the Higgs field and remain massless. Just as an accelerating body `carries' a flow and acquires an added mass, it is as if the $W$ boson `carries' the Goldstone mode and becomes massive. Remarkably, the elusive Higgs particle has a simple interpretation in this analogy. Indeed, the Higgs particle which is the longest wavelength mode of oscillation of the Higgs field, may be thought of as analogous to a long-wavelength wave in the fluid around an accelerated body. More generally, quantum fluctuations around the scalar vacuum expectation value are analogous to density fluctuations (e.g. sound waves) around constant density flow. In fact, the semi-classical `loop expansion' in powers of Planck's constant ($\hbar$) that accounts for quantum fluctuations is analogous to an expansion in powers of the Mach number that is used to describe effects of compressibility around the limit of incompressible (constant density) flow. We conclude by summarizing the analogy between the Higgs mechanism and the added mass effect in Table \ref{t:HAM}.

\noindent {\bf Acknowledgements:}
The authors would like to thank A. Thyagaraja for enlightening discussions on the added mass effect. Thanks are also due to an anonymous referee for useful comments and suggestions.

\appendix
\section{Bernoulli equation for unsteady potential flow}
\label{a:bernoulli-unsteady-potential-flow}

In fluid mechanics we often encounter the Bernoulli principle for steady flow\cite{ref-feynman-vol-2, ref-deshpande}. A flow is steady if the fluid velocity at any location is independent of time. Bernoulli's equation states that $p + \half \rho v^2 + \rho g h$ is a constant along streamlines (path followed by a test particle in steady flow). Here $\rho$ is the constant density of the fluid, $g$ the acceleration due to gravity and $h$ the height of any point on the streamline. It follows that a rise in pressure is often associated with a drop in flow speed or potential energy.
		
In our discussion of the added mass effect, we require a generalization of the Bernoulli equation to the \emph{unsteady} potential flow around an accelerated body. To derive this equation, we will first obtain Euler's equation of fluid mechanics which is Newton's second law for a small parcel of fluid of volume $\del V$. We begin by noting that the change in velocity of such a `fluid element' over a small time $dt$ as it moves from position $\bfr$ to $\bfr + d\bfr$ is
\beq
	d\bfv = \bfv(\bfr + d\bfr, t + dt) - \bfv(\bfr, t) \approx \dd{\bfv}{t} dt + (d\bfr \cdot \grad) \bfv.
\eeq
Dividing by the small time, letting $dt \to 0$ and observing that $\DD{\bfr}{t} = \bfv$, we obtain the instantaneous acceleration of a fluid element of mass $\rho \,\del V$:
\beq
	\DD{\bfv}{t} \equiv \dd{\bfv}{t} + (\bfv \cdot \grad) \bfv.
\eeq
$\DD{\bfv}{t}$ is called the material derivative, it differs from the partial derivative by the quadratically non-linear advection term $(\bfv \cdot \grad)\bfv$. In the absence of gravity, the only force on this fluid element is due to the pressure exerted by the surrounding fluid. This force acts across the surface $\partial (\del V)$ of the element and is given by:
\beq
	\bfF_{\rm surface} = - \int_{\partial (\del V)} p \bfhatn \,dS = -\int_{\del V} \grad p\, dV \approx -\grad p \,\del V.
\eeq
We have used a corollary\footnote{Gauss' divergence theorem relates the integral over a volume $V$ of the divergence of a vector field $\bfq$ to the flux of the vector field across its bounding surface $\partial V$:
\beq
	\int_V \grad \cdot \bfq\, dV = \int_{\partial V} \bfq \cdot {\bf \hat n} \,dA.
\eeq
Taking $\bfq$ to have a constant direction $\bfq = \varphi(\bfr) {\bf c}$, where ${\bf c}$ is an arbitrary constant vector and $\varphi(\bfr)$ is any scalar function, we obtain a corollary of the divergence theorem:
\beq
	{\bf c} \cdot \int_V \grad \varphi\, dV = {\bf c} \cdot \left( \int_{\partial V} \varphi {\bf \hat n}\, dA \right) \quad \text{or} \quad \int_V \grad \varphi\, dV = \int_{\partial V} \varphi {\bf \hat n}\, dA.
\eeq

} of Gauss' divergence theorem to convert the surface integral to a volume integral and taken $\grad p$ to be constant over the small volume $\del V$. The minus sign is because $\bfhatn$ is the outward-pointing normal to the surface, while $\bfF_{\rm surface}$ is the compressional force on the element due to its neighbours. Newton's second law of motion for the fluid element therefore reads
\beq
	(\rho \,\del V)\,\DD{\bfv}{t} = \grad p \,\del V
	\quad \text{or} \quad \dd{\bfv}{t} + (\bfv \cdot \grad) \bfv = -\frac{\grad p}{\rho}.\label{eqn-euler}
\eeq
This is Leonhard Euler's celebrated equation (1757) for flow of an inviscid fluid. Using the vector identity $(\bfv\cdot \grad) \bfv = \grad \left( \half v^2 \right) + (\grad \times \bfv) \times \bfv$ \cite{ref-feynman-vol-2} and observing that for potential flow $\grad \times \bfv = \grad \times (\grad \phi) = 0$, we may rewrite Euler's equation for constant density as
\beq
\dd{\bfv}{t} + \grad \left(\half v^2 \right) = -\frac{\grad p}{\rho} \quad \text{or} \quad \grad \left( p + \rho \dd{\phi}{t} + \half \rho v^2 \right) = 0.
\eeq
Thus we arrive at the unsteady Bernoulli equation for potential flow where the Bernoulli constant $B(t)$ is independent of location but could depend on time:
\beq
p + \rho \dd{\phi}{t} + \half \rho v^2 = B(t).\label{eqn-unsteady-bernoulli}
\eeq
It may be interpreted as an equation for the evolution of the velocity potential $\phi$. It may also be used to eliminate the pressure $p$ in favor of the velocity potential, as we do in Section \ref{sec-derivation-subsec-1}. In contrast to Bernoulli's equation for steady flow, (\ref{eqn-unsteady-bernoulli}) holds throughout the fluid and is not restricted to streamlines.


\end{document}